# Effect of Cultural Factors on Students of Pakistan

Nazia Hameed, Maqbool ud din shaikh, Fozia Hameed

***Abstract***— Pakistan as culturally diverse country possesses wide range of cultural factors. These cultural factors affect the living style, traditions, values and norms as well as the education. As e-learning is global mode of education so learners from different backgrounds enrolled in learning management system. Diversity of culture and learning styles should keep under considerations while designing e-learning environment. In this research work e-learning cultural factors highlighted by bently et. al are incorporated to find the learner level. After assignment of level, learner respective Learning Management System(LMS) is allocated to learner. The focus is to concentrate on cultural factor in e-learning system. Furthermore; a prototype of the proposed system is implemented for the validation of proposed architecture.

## I. INTRODUCTION

The most promising outcome of the Internet and information technology is e-learning. E-learning has made rapid development and has become a global mode of education. It is an umbrella term that describes learning done at a computer, online or offline. E-learning is the process of delivering knowledge to the learner with the help of computers, the Internet, Intranet, the Web, from the hard drive, CD or DVD of the learner computer. In e-learning there are many issues need to be further investigated. One of the issues that need to be explored is cultural differences [1],[2],[7]. It is an umbrella term that includes living life style, education system, place, gender, race, history, nationality, language, sexual orientation, religious beliefs, ethnicity and aesthetics.

Pakistan is a developing country occupying a vital geographical position. Different regions of Pakistan are culturally different from each others. In Pakistan, there are many educational institutes having different educational environment. Student enrolled in e-learning system belongs to different regions and different educational institutes having different culture values. Their education system, medium of instruction, course contents etc. are different. Students of backward areas don't have good exposure of information and computer technology (ICT). Most of existing e-learning systems do not consider these differences due to which learners face many problems during their study. Cultural differences are major problem which effect e-learners [1],[3],[4],[5],[6],[7].Most of current e-learning system does not consider cultural diversity of learners.

In this paper, authors designed and proposed an e- learning system that caters the cultural difference issue of learners. Furthermore, authors have developed a prototype of proposed system.

The rest of the paper is organized as follows. Section II illustrates the literature review; Section III describes the proposed architecture, section IV presents system implementation and section V shows results and discussions.

## II. LITERATURE REVIEW

### A. e-learning

According to Badrul H. Khan, "E-learning can be viewed as an innovative approach for delivering well-designed, learner-centered, interactive, and facilitated learning environment to anyone, anyplace, anytime by utilizing the attributes and resources of various digital technologies along with other forms of learning materials suited for open, flexible, and distributed learning environment" [6]. E-learning is also defined as "e-learning is the use of technology to enable people to learn anytime and anywhere. E-learning can include training, the delivery of just-in-time information and guidance from experts" [7].

### B. Mode of e-learning

There are two modes of e- learning, i.e. synchronous and asynchronous. In synchronous mode, classes take place in a classroom in real-time. Communication between instructors and learners through teleconferencing or a chat room are examples of this mode. In asynchronous mode, learners can access educational material in their convenient time anywhere. In synchronous mode, learning takes place in real time but in asynchronous, it is not take place in real time.   A brief description of these modes is presented in [12],[13].

### C. Modalities of e-learning Activities

According to Som Naidu, there are four types of e-learning activities i.e. individualized self-paced e-learning online, individualized self-paced e-learning offline, group-based e-

This work was supported by COMSATS Institute of Information Technology Islamabad Campus.
Nazia hameed is with the COMSATS INSTITUTE of Information Technology Islamabad Pakistan, (Phone: +92 312 5062757 e-mail: nazia_hameed@comsats.edu.pk).
Maqbool ud din shaikh is with Preston University Islamabas Pakistan (email: maqboolshaikh@preston.edu.pk )
Fozia hameed is with the KING KHALID UNIVERSITY Saudi Arabia, ( e-mail: fozi_aug12@yahoo.com ).

learning synchronously; and group-based e-learning asynchronously [14]. In individualized self-paced e-learning online; learners use the internet for accessing learning resources. In individualized self-paced e-learning offline, learners access learning resources without the Internet. In group-based e-learning synchronously, groups of learners are working together in real time via the Internet. In group-based e-learning asynchronously, groups of learners are working over the Internet but not in real time[10], [14].

*D. e- learning Models*

Several e-learning models are proposed by different researchers [6],[9],[10],[14]. Some models emphasis on the interactions between the teacher, student and the content [5], [8] while others focus on embedding tacit knowledge in pedagogical model to enhance e-learning [13].

Khan proposed P3 model for e-learning in [6].The P3 model is divided into two phases; (1) content development and (2) delivery and maintenance. Content development involves planning, design, development and evaluation of e-learning content and resources. Implementation of online course offerings, monitoring and updation of e-learning environment are included in delivery and maintenance phase.

Terry et. al. in [9] proposed a model of e-learning in which different types of interactions are illustrated. Two major human characters of e-learning i.e. teachers and student are represented in the model. It represents interaction between the teacher, student and content. Learners can directly interact with the content as the e-learning material can easily be access using the Internet.

Rizwana et. al. in [14] proposed a structure for embedding tacit knowledge in pedagogical model. Her proposed pedagogical framework includes following three phases; content organization, quality assessment and content delivery phase. In the contents organization phase, the contents used for learning are collected and organized based on the requirements put forward by the industries. Quality assessment phase analyzes and assesses the contents developed and gathered based on theoretical and experimental methods. In contents delivery phase, the learner is assessed to find out their aptitude and level.

Another architecture of e-learning system is proposed in [10] that incorporates industrial and academics needs and requirements both. Contents are developed on the basis of these needs and requirements. To enhance the quality of e-learning system, quality assessment and assurance processes are added into this architecture. These processes will insure the quality of knowledge delivered to the learner.

*E. Culture Difference in e-learning*

Culture can be defined as "the beliefs, value systems, norms, mores, myths, and structural elements of a given organization, tribe, or society" [16]. Culture refers to the norms, beliefs of the groups of individuals living in the society. Culture affects the learner's behaviors of online education and these effects must be taken into consideration to make e-learning more effective and practical. As shown in Fig 1, the cultural environment affects online learning environments in two aspects i.e. design and use.

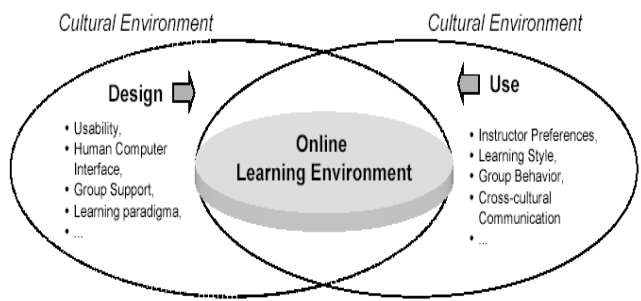

Figure 1: Effect of Cultural Environment on online learning Environments

*F. Effects of Culture on the Design and Development of e-learning System*

E-learning system is build taking into consideration, needs and the customs of learners. Without keeping these considerations the system cannot be effective and thus unable to attain the required results. To analyze the differences, the communication disciplines involved are computer-mediated communication, cross-cultural communication, cultural context interaction mode and communication form [17] as shown in Fig 2.

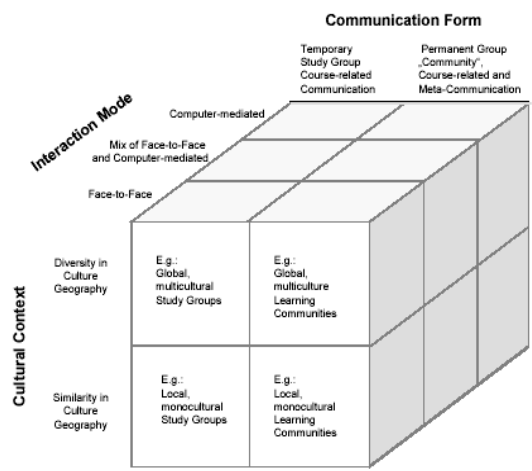

Figure 2: Cultural Environments

With the advent of new technologies in distributed learning, e-learning courses can be accessed throughout the global world. While designing e-learning environments the diversity of culture and learning styles should kept under considerations in order to enhance learning for all. One of the difficult issues in e-learning is cross-cultural communication [18].Gender also affect the learning capabilities of the learners [19].

Eight educational value differentials are highlighted by Bentley et. al. in [22] to understand the cultural issues. These issues are language differential, educational culture difference, technical infrastructure differential, local versus global differential, learning style differential, reasoning pattern differential, high- and low-context differential and social context differential. The authors not only incorporate the language and technology but also learning styles and local and global context.

### III. PROPOSED ARCHITECTURE

The proposed architecture is discussed below and shown in Fig 3.

#### A. e-learning

In registration process, student registers a course through e-learning system. In first step student is asked for signup. During signup student enter two types of data; personal and cultural. Student data is stored in student data repository which is divided into three sub repositories i.e. personal repository, cultural repository and feedback repository.

• Personal Repository

Personal information of students is stored in the personal database.

• Cultural Repository

To cater cultural diversity, student's cultural and educational information regarding is required. After gathering personal information student enter his educational and culture information. Cultural educational background includes the school type i.e. government school or private school, medium of instruction of primary secondary and master level, course contents of previous study, area from which student belong, computer skills, school environment, economic background etc. This information is stored into educational and cultural repository. When student entered personal, educational and cultural information he is successfully registered in the system.

• Feed Back Repository

Students feed backs are stored in this repository for future decision making.

#### B. Aptitude Test

Aptitude test is used to judge the student knowledge level. After signing in student take the aptitude test. Quality is a major concern in designing of the aptitude test. In our system Quality Manager assure that the questions designed for aptitude test are according to educational standards. Aptitude test consists of four portions; English, Mathematical Reasoning, Computer and Intelligence Quotient. Students registering for e-learning should know basics of computer and technology. Therefore, to judge the student knowledge about the computer portion related to computer questions is included.

Each portion consists of ten questions and each question carry one point. Aptitude test starts with English portion. When English portion is successfully completed, students solve mathematical reasoning, computer and intelligence quotient portions respectively.

#### C. Aptitude Test Evaluation

When aptitude test is successfully submitted, the score of English ($S_E$), Mathematical Reasoning ($S_{MR}$), Computer ($S_C$) and Intelligence Quotient ($S_{IQ}$) is evaluated individually.

$$Total = S_E + S_{MR} + S_C + S_{IQ} \ldots\ldots\ldots (1)$$

Total score of aptitude test is calculated using equation 1. After evaluating total, percentage is calculated. Total score and percentage of a student's aptitude test is then stored into in the student repository.

#### D. Inference Engine

Inference engine check the educational background of the student and calculate the reference value ($refvalue_{student}$) of the student. $refvalue_{student}$ of students depends on the cultural and educational information. $refvalue_{student}$ is based on following factors:

• Learning language (Medium of Instruction)

Medium of instruction is a key factor that should be considered in e-learning system. Students who have English language as medium of instruction perform better in education as well as got better jobs then the students who have local language as the medium of instruction [20]. Hengsadeekul et. al in [21] emphasize the importance of English and conclude that for better intercultural communication English should be better understand.

• Computer Knowledge

An important factor in calculation of $refvalue_{student}$ is student's knowledge about computer. Marzano et. al. in [23] stressed the importance of having background knowledge and concluded that insufficient background knowledge causes lower achievement in learners.

• Course Contents (Local or International)

Another Important factor is course contents of the student's previous study. Contents of the course and the language affect a lot on the learner knowledge [24].

• Percentage Marks of Aptitude Test

Other factor considered for $refvalue_{student}$ is student's percentage marks of aptitude test. Levels are assigned to the students and level depend on $refvalue_{student}$ and Total.

There are three levels in which students are categorized listed below:

• Beginner
• Intermediate
• Skilled

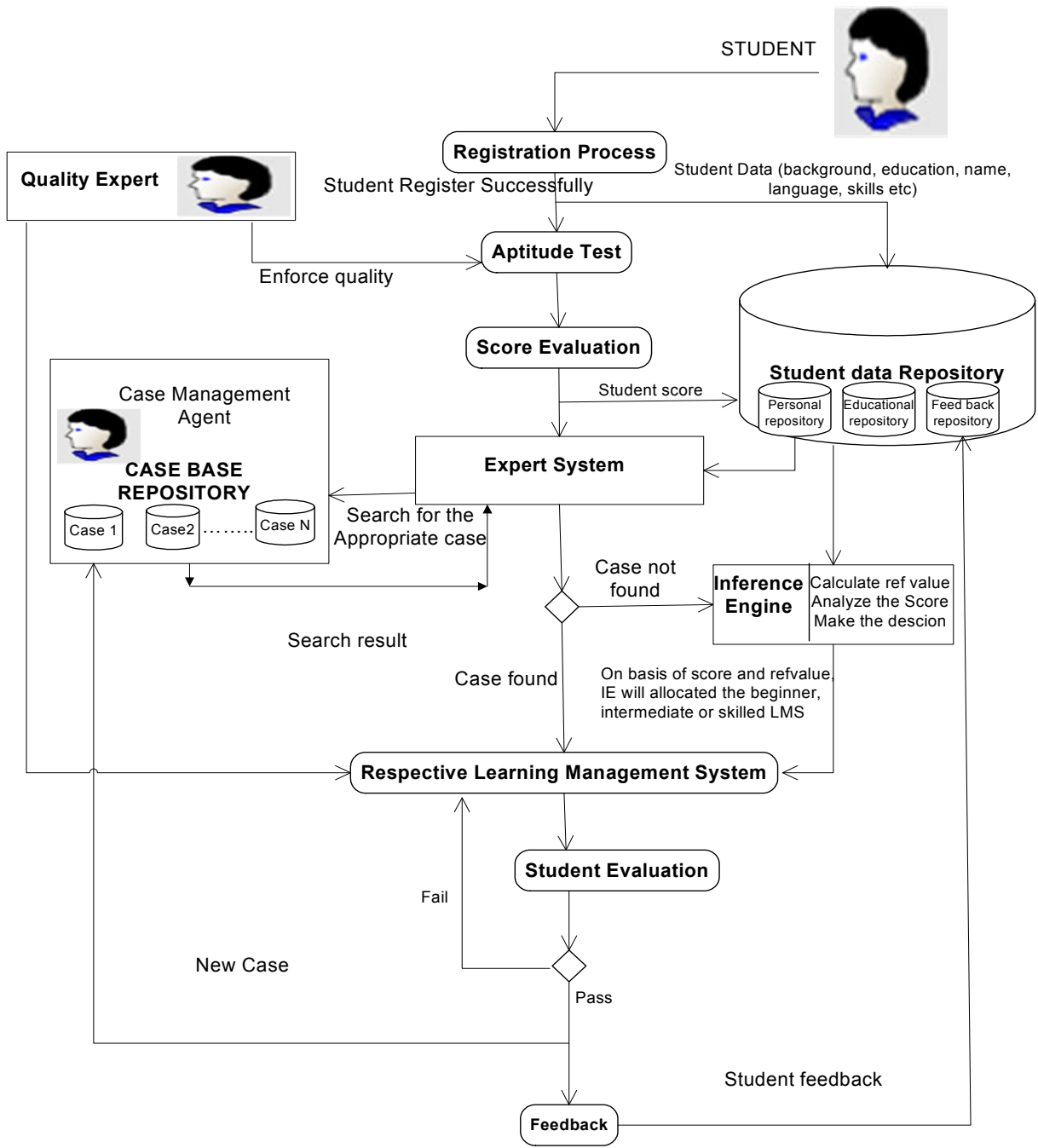

Figure 3: Proposed Architecture

The assignment of levels is summarized in the pseudo code below:

IF ((( $Average_{refvalue}=7$ ) && ( $\%age>=60$ )) OR (( $Average_{refvalue}=6$ ) && ( $\%age>=70$ )) OR (( $Average_{refvalue}=5$ ) && ( $\%age>=80$ )) OR (( $Average_{refvalue}=4$ ) && ( $\%age>=85$ )) OR (( $Average_{refvalue}=3$ ) && ( $\%age>=90$ )))

{
Level = Skilled
}
Else IF ((( $Average_{refvalue}=7$ ) && (( $\%age<60$ ) && ( $\%age>=50$ )) OR (( $Average_{refvalue}=6$ ) && (( $\%age<60$ ) && ( $\%age>=50$ ))) OR (( $Average_{refvalue}=5$ ) && (( $\%age<75$ ) && ( $\%age>=60$ ))) OR (( $Average_{refvalue}=4$ ) && (( $\%age<85$ ) && ( $\%age>=70$ ))) OR (( $Average_{refvalue}=3$ )&& (( $\%age<95$ ) && ( $\%age>=80$ )))
{
Level = Intermediate
}
Else IF ((( $Average_{refvalue}=7$ ) && (( $\%age<50$ ) && ( $\%age>=40$ )) OR (( $Average_{refvalue}=6$ ) && (( $\%age<50$ ) && ( $\%age>=40$ ))) OR (( $Average_{refvalue}=5$ ) && (( $\%age<60$ ) && ( $\%age>=40$ ))) OR (( $Average_{refvalue}=4$ ) && (( $\%age<70$ ) && ( $\%age>=40$ ))) OR (( $Average_{refvalue}=3$ )&& (( $\%age<80$ ) && ( $\%age>=40$ )))
{
Level = Beginners
}
Else IF ( $\%age<40$ )
{
    Then not eligible for the degree
}
End IF

*E. Inference Engine*

After the levels have been assigned to the students, student is allocated to the respective learning management system (LMS) as shown in Fig 4.LMS is defined as "LMSs facilitate learning by providing a centralized location were learning material reside and through integrated tools that enable various teaching and learning activities such as communication and collaboration with peers and lecturers, self-assessment and progress tracking" [25].

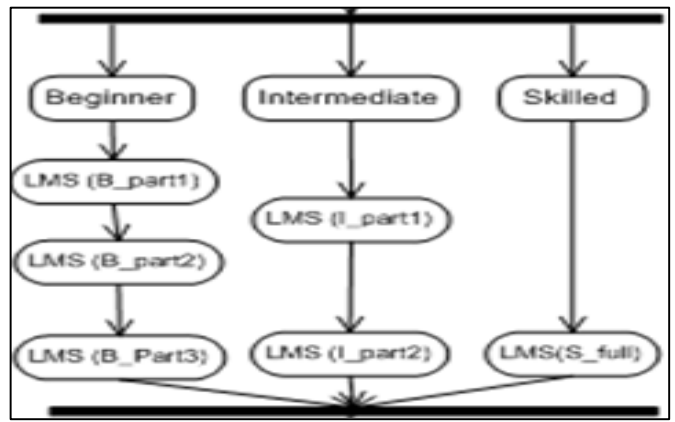

Figure 4: LMS for Beginners, Intermediate and Skilled

*F. Student Evaluation*

Students are evaluated after they successfully complete the course. During their course they are evaluated frequently with assignments and quizzes. If the students fail in evaluation they again take the course.

*G. Case base Reasoning*

Case base reasoning is an additional feature in the proposed architecture. Every student who successfully passes the course is stored as new case in case base repository with all the relevant data. When new student is registered, his educational culture background is matched with the cases stored in case base. If the case matches with existing case, the LMS of match case is assigned to newly registered students.

*H. Feedback*

Feedback from students should be considered in to e-learning systems. Feedback form is evaluated from students who successfully pass the course. The suggestions are stored in the feedback repository. These suggestions are helpful for making future improvements in the system.

IV. SYSTEM IMPLEMENTATION

In the working prototype, there are two main interfaces (1) admin end and (2) user end.

*A. Admin End*

Admin end is created to facilitate the administrator of the system. Using admin panel, administrator can manage aptitude test. Administrator can easily add, modify and delete questions in the test. Fig 5,6,7,8 shows main admin panel window. When admin successfully login into the system, administrator can perform function like managing test, system and students.

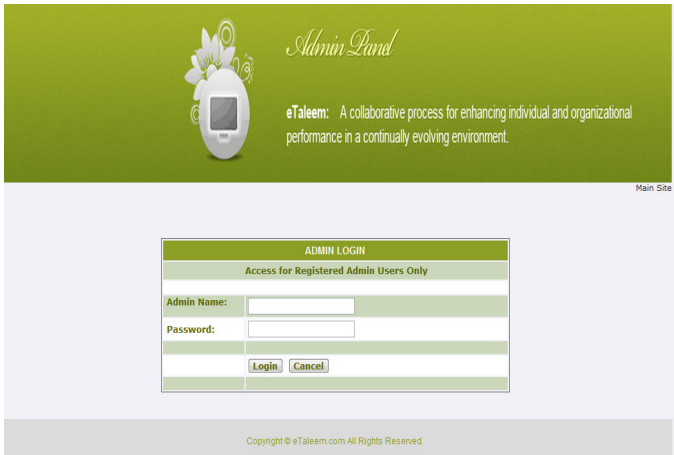

Figure 5: Admin Panel (Login )

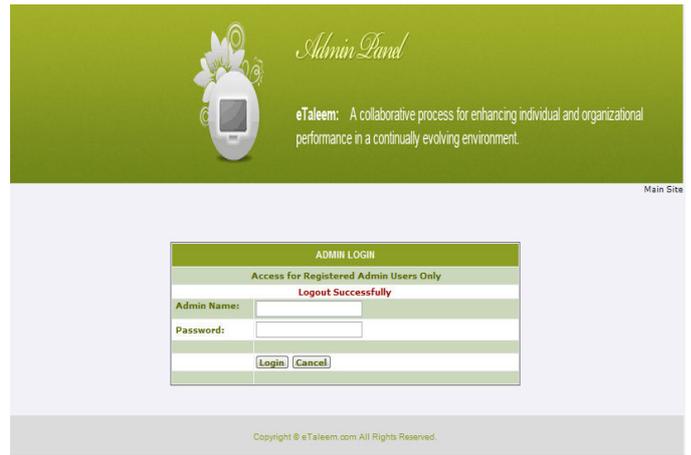

Figure 8: admin panel(Log out )

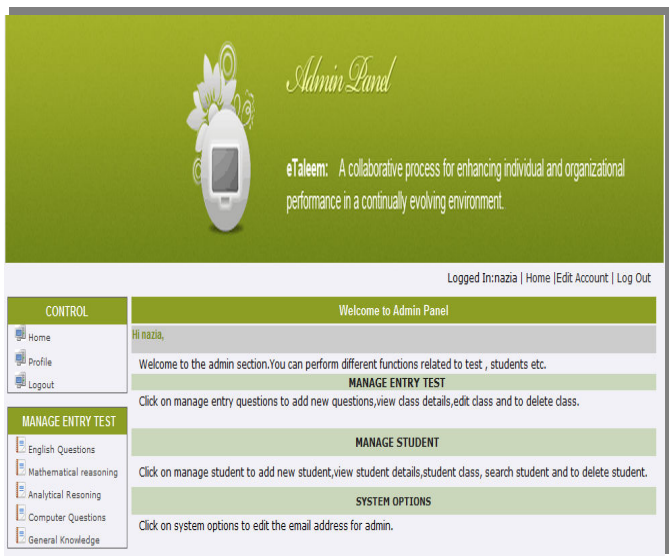

Figure 6: Admin panel Home page

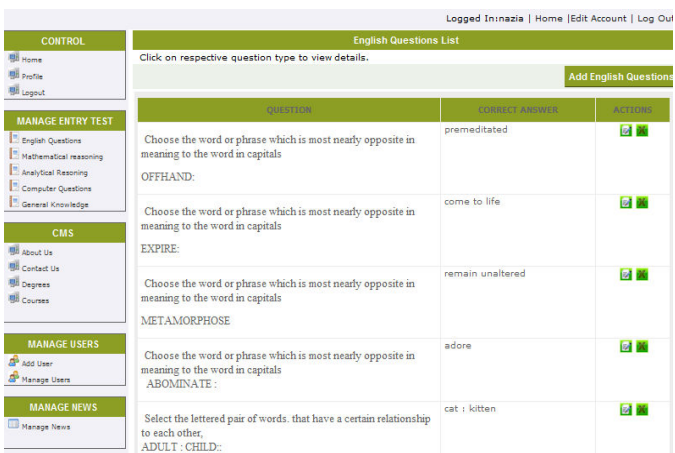

Figure 7: Admin panel (Manage Test )

*B. User End*

User end is used by students for registration and to take aptitude test. Fig. 9 shows the front end. Users first have to register themselves before taking any course. During registration student will enter its educational and cultural information as shown in Fig. 10.

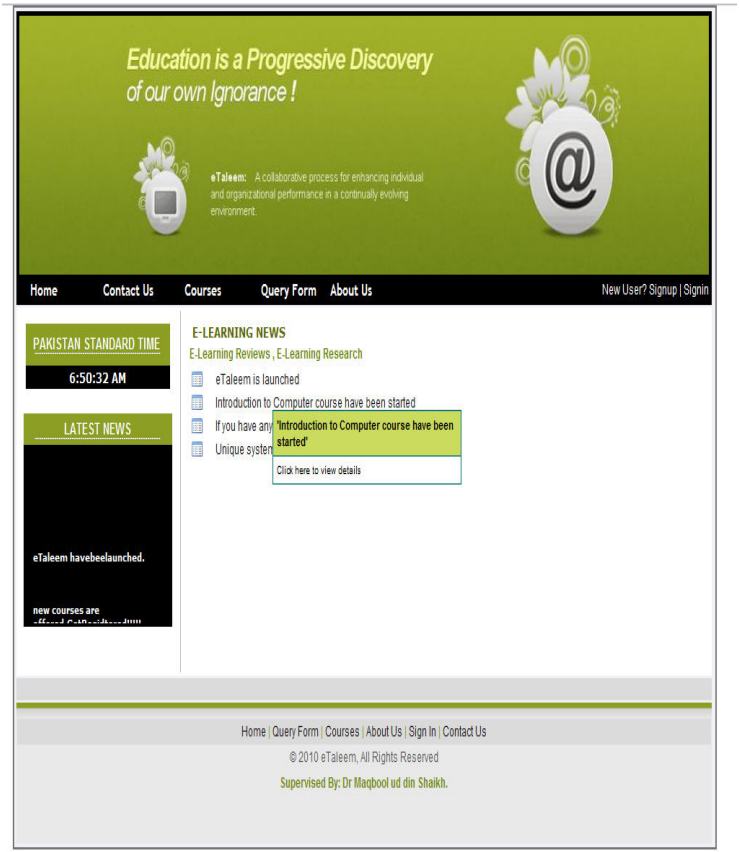

Figure 9: User End

Figure 10 (a): Registration Process

Figure 10 (b): Registration Process

After successful registration student will sign in to take the test. Login screen is shown in Fig 11.

Figure 11: Login

When student click on start Aptitude test button, test starts.

Figure 12: Test Start screen

Test consists of different portions. A portion of English test is shown in Fig 13.

Figure 13: English Test

Similarly student attempts computer and intelligent quotient part. On the completion of Aptitude rest, student level is calculated based upon the reference value and the marks obtain by the student. Student level is displayed on the screen as shown in Fig 14. After assigning the level, student will directed to respective LMS.

Figure 14: Student level

V. RESULTS

The prototype is evaluated from different users. Authors evaluated the prototype from students of Islamabad Model College for girls, (IMCG) F6-2, Islamabad and COMSATS University, of Science and Technology, Islamabad. 54% students are male and 46% are female as shown in Figure 15.

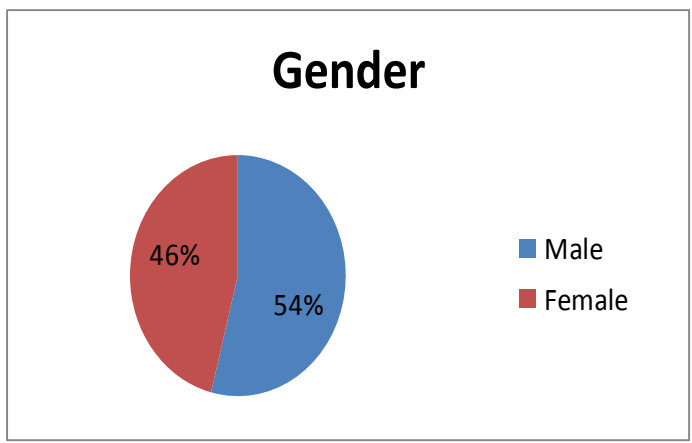

Figure 15: Research Result (Population)

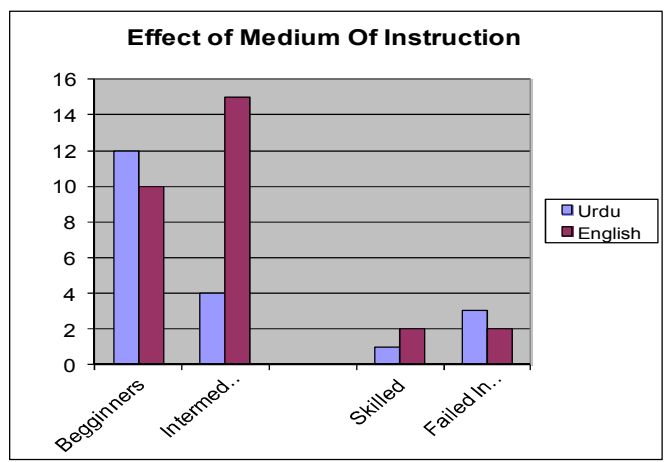

Figure 17: Effect of Medium of Instruction

Students categorized into different levels as shown in Fig 16 prove that the educational factors affect e-learning. Therefore the student's cultural diversity should be considered while offering e-learning programs. Majority of students having Urdu language as their medium of instruction falls in the beginner level, therefore the conclusion develop is that medium of instruction affect a lot on student learning as shown in Figure 17.

Acknowledgment

First and foremost, my utmost gratitude to COMSATS INSTITUTE OF INFORMATION TECHNOLOGY which support me in this work. Thanks to Dr. Maqbool ud din sheikh , Professor in the Preston university guided me to complete this research work. Last but not the least I would like to thanks WSEAS team and its reviewers who review the paper and guided us to improve the research work.

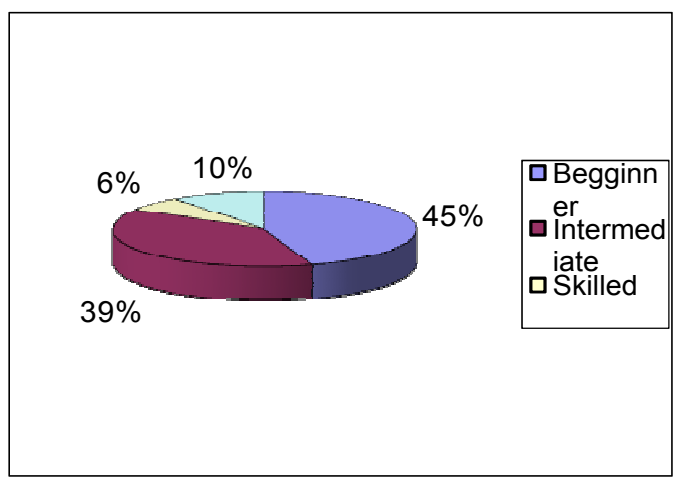

Figure 16: Effects of Education Factors on e-learning

Results gathered shows that the students who have studied international course contents have better score results then those who studied the local course contents, which concludes that course contents also affects the learners.

Students who don't have exposure to computer face many problems during e-learning therefore Human Computer Interaction and its principles should be considered while designing systems for students having different educational culture.

REFERENCES

[1] C. Belisle,( 2008, February) "E-learning and intercultural dimensions of learning theories and teaching models" eLearning Papers, available at http://www.elearningeuropa.info/files/media/media14903.pdf, ISSN: 1887-1542

[2] A. Edmundson, Globalized e-learning Cultural Challenges , by Information Science Publishing, USA 2007.

[3] M. Milani, Cultural Impact on Online Education Quality Perception. The Electronic Journal of e-Learning. Volume 6 Issue 2, April 2008, pp 99 – 182

[4] A.Nawaz and G. M. Kundi, "Predictor of e-learning development and use practices in higher education institutions (HEIs) of NWF Pakistan", Journal of Science and Technology Education Research (JSTER) Vol. 1, August 2010.

[5] DR. Jaflah Al-Ammari ,Ms. S. Hamad , "Factors influencing the adoption of e-learning at UOB", Second International Conference and Exhibition on "e-Learning and Quality Education and Training, 28 -30 April 2008, Bahrain

[6] B.H. Khan, "Managing E-learning Strategies: Design, Delivery, Implementation and Evaluation", Information science Publishing, United States of America, 2005.

[7] D.Chiribuca, I. Pah, D. Hunyadi, "Social and Cultural Challenges of the New Communication Technology Used in Education Oriented Activities ", 7th WSEAS International Conference on EDUCATION and EDUCATIONAL TECHNOLOGY (EDU'08) Venice, Italy, November 21-23, 2008.

[8]  E-learning    consulting,    available    at    http://www.e-learningconsulting.com/consulting/what/e-learning.html, Accessed date 5 July, 2010

[9]  T. Anderson, F. Elloumi, 2nd Edition, Theory and Practice of Online Learning, AU Press, Canada, 2008.

[10]  M. udin Shaikh, Azra Shamim, "Proposed Architecture for e - Learning System Incorporating Industrial and Academic Requirements", 5th International Conference on Information Technology Jordan, 11-13 May, 2011.

[11]  Hoskins, S. L., & van Hooff, J. C. (2005) " Motivation and ability: which students use online learning andwhat influence does it have on their achievement?", British Journal of Educational Technology, 36, 2 p177-192

[12]  S. Hrastinski, "A study of asynchronous and synchronous e-learning methods discovered that each supports different purposes",by EDUCAUSE Quarterly, vol. 31, no. 4 (October–December 2008).

[13]  K. Hyder, A Kwinn, R. Miazga, and M. M. Tony Bates, "The eLearning Guild's Handbook on Synchronous e-Learning", by The eLearning Guild, 2007.

[14]  S. Naidu, "E learning a Guidebook of Principles, Procedures and Practices"( 2nd Revised Edition, 2006), by CEMCA, New Delhi

[15]  R. Irfan,M. Uddin Shaikh, "Framework for Embedding Tacit Knowledge in Pedagogical Model to Enhance E-learning", In proceeding of 2nd International conference on New technologies, Mobility and security, Tangier , 5-7 Nov 2008, pp 1-5

[16]  R. T. Watson, T. Hua Ho and K. S Raman, "Culture: a fourth dimension of group support systems", Communications of the ACM, Oct. 1994.

[17]  S. Seufert, "Handbook of Information Technologies for Education and Training", Springer- Verlag Berlin Heidelberg New York.

[18]  Sanchez  and Gunawardena, "Understanding and supporting the culturally diverse distance learner", by Atwood Publishing Madison, 1998.

[19]  S. Hassan, N. Ismail, K. Ghazali, A. Shahira, A. Samad, "Gender Comparison on the Factors Affecting Students' Learning Styles", 10th WSEAS International Conference on Education  and Educational Technology (EDU '11)

[20]  Z. Zaaba,  I.N. Anthony Aning, H. Gunggut, F Ibrahim Mahmood, K. Umemoto , "English as a Medium of Instruction in the Public Higher Education Institution: A Case Study of Language-in-Education Policy in Malaysia", 9th WSEAS International Conference on Education  and Educational  Technology Iwate Prefectural University, Japan, October 4-6, 2010

[21]  C. Hengsadeekul,  T. Hengsadeekul ail, R.Koul, S. Kaewkuekool, "English as a Medium of Instruction in Thai Universities: A Review of Literature", 9th WSEAS International Conference on Education  and Educational  Technology Iwate Prefectural University, Japan, October 4-6, 2010

[22]  Bentley, P. H., M. V. Tinney, and B. H. Chia, "Intercultural internet-based learning: Know your audience and what it values", Educational Technology Research & Development, 2005.

[23]  R. J Marzano, "Building Background knowledge for academic achievement", Association for Supervision and Curriculum Development, 2004.

[24]  G. Shaheen, S. Rehman, N. Bajwa , Umbreen Ishfaq , M Naseer Ud Din , "A Comparative Study of Pakistani Expert's Perception About the Examination System Of GCE (A Level) and Fsc in Chemistry", International Journal of Academics Research, Vol. 2. No. 6. November, 2010.

[25]  Hatziapostolou, T and Paraskakis, I. (2010), "Enhancing the Impact of Formative Feedback on Student Learning".

**Nazia hameed** is currently appointed as lecturer in COMSATS INSTITUTE of Information Technology Islamabad . She completed her MS(CS) from COMSATS INSTITUTE of Information Technology Islamabad in March 2011. Her research Interests are e-learning, content base image retrieval.

**Maqbool Ud din Shaikh** is currently appointed as professor  in Preston University Islamabad Pakistan. He completed his Phd from The Liverpool University, U.K. in 1987. He has more than 20 year experience in educational field and supervised many MS and Phd students. His research interest are e-learning, e-government, data mining, web mining , software engineering.

**Fozia hameed** is currently appointed as lecturer in KING KHALID University Saudi Arabia . She completed her MS(CS) from COMSATS INSTITUTE of Information Technology Islamabad  in March 2008. Her research interests are e-activities, wireless networks.